# PRIVACY CHALLENGES IN IMAGE PROCESSING APPLICATIONS

Maneesha

BITS Pilani Dubai Campus, Dubai,
United Arab Emirates

maneesha@dubai.bits-pilani.ac.in

Bharat Gupta

BITS Pilani Dubai Campus, Dubai,
United Arab Emirates

f20220072@dubai.bits-pilani.ac.in

Rishabh Sethi

BITS Pilani Dubai Campus, Dubai,
United Arab Emirates

f20230307@dubai.bits-pilani.ac.in

Charvi Adita Das

BITS Pilani Dubai Campus, Dubai,
United Arab Emirates

f20240224@dubai.bits-pilani.ac.in

## Abstract

As image processing systems proliferate, privacy concerns intensify given the sensitive personal information contained in images. This paper examines privacy challenges in image processing and surveys emerging privacy-preserving techniques including differential privacy, secure multiparty computation, homomorphic encryption, and anonymization. Key applications with heightened privacy risks include healthcare, where medical images contain patient health data, and surveillance systems that can enable unwarranted tracking. Differential privacy offers rigorous privacy guarantees by injecting controlled noise, while MPC facilitates collaborative analytics without exposing raw data inputs. Homomorphic encryption enables computations on encrypted data and anonymization directly removes identifying elements. However, balancing privacy protections and utility remains an open challenge. Promising future directions identified include quantum-resilient cryptography, federated learning, dedicated hardware, and conceptual innovations like privacy by design. Ultimately, a holistic effort combining technological innovations, ethical considerations, and policy frameworks is necessary to uphold the fundamental right to privacy as image processing capabilities continue advancing rapidly.

## I. INTRODUCTION

Imaging has become ubiquitous in the modern world, being used everywhere from social media to surveillance to healthcare. However, such extensive collection, storage, and processing of image data raises serious privacy concerns that must be addressed. Images contain deeply emotional information about individuals - their physical appearance, their actions, their company, their locations, and even their emotional state. The unauthorized use of a person's images is a serious breach of privacy. At the same time, image analytics drives innovation in security, medicine, etc. - so balancing privacy and processing creates unique challenges. Many parts of the image processing machine create privacy vulnerabilities. Personal data may be collected without consent through surveillance and recording during image recording. Image data storage carries the risk of exposure through hacking and data breaches (Seh et al. 2020). In addition, some automated images can reflect individual characteristics. Face recognition analysis can match images with human identities without permission. Medical imaging analysis can reveal patients' health status. An algorithm that operates in a learning manner may also recall the private information of the training datasets. As imaging capability skyrockets, the vulnerability to privacy infringements expands exponentially. Approaches that ensure enough security in imaging need to be implemented to protect privacy. Stipulating that technological solutions are required to safeguard privacy, the following ones can be given encryption measures, access control means as well as algorithms for privacy. Policies in image data allowed uses also have an impact on the same.

Thus, conductive multi-faceted strategies are required to resource the single privacy in terms of moving advanced image data collection environment. Imagining modalities is an essential topic to discuss several privacy issues in this paper. More specifically, the paper provides insight into challenges of privacy that are associated with capturing receptive data during their storage and processing such as risks of the emergence of these incidences, discusses various methods and approaches and also addresses privacy risk management components to be used to address these problems. Conservation becomes the top focus and grounds for further learning to understand all the means used to achieve this objective.

## II. Safeguarding Sensitive Information In Surveillance And Healthcare

In applications such as research and healthcare, the protection of sensitive information is critical. When it comes to data protection, the research and healthcare application is key in ensuring that your data remains private. Privacy controls are needed to preserve the confidentiality of individuals' identities and personal data when such persons become subjects of surveillance images or videos (Shokri et al., 2017). Camera works are everywhere- starting from security checkpoints to smart city surveillance systems. A person's civil liberties are constrained by excessive external supervision of such a person's activities, and the prevention of crime is thereby facilitated as well as law enforcement. The complex video analytics involved in the advanced technologies further raises the risks of philosophical quantification by automating the detection and visualization of objects. As a measure of privacy, effective surveillance protects both by enabling the use of an appropriate system without compromising security objectives. User permission and restricted accessibility to encrypted views that are hidden from the public, prevent users with no permissions from viewing the data. Blurring, pixelation, and masking do not name faces to recognition as it fail at identifying the persons correctly in these situations. It is possible to conduct the search question on encrypted data due to secure multiparty computation. The improved simulations that are closely related to computer vision make privacy more efficient. In total, establishing a balance between searching and confidentiality needs a combined –system makeup of data control, algorithms created by software pieces, and engineering makeup.

For health care, the images for CT scan pictures, X-rays, etc (Seh et al. 2020). Such are medical imaging of great help in diagnosis and treatment planning. But it is packed with material that contains very sensitive patient health information. Informing people that their health information is being shared with others can discourage them from seeking care or betrayal of trust. However, the pictures are to be kept confidential, but they must still have viewership from authorized users. As for Homomorphic encryption, it correlates with the function that performs computations immediately in encrypted form without decryption. Differences in confidentiality, however, mean this noise of the limits of revelation on the results of a study helps to ensure confidentiality. Access controls and data separation prevent visibility. Together, these cryptographic and algorithmic safeguards support patient privacy by providing important benefits for quality healthcare.

*A. Advanced Privacy-Preserving Techniques for Image Analysis*

In addition to basic techniques such as basic techniques such as access and encryption, advanced privacy-preserving computing techniques perform more granular, yet privacy-sensitive analysis of sensitive image data ban These techniques allow us to extract a combination of useful insights from images without compromising individual privacy. The following sections will focus on specific privacy protection measures, including:

1) *Differential Privacy in Image Processing*

Differential privacy is a cryptographic technique (**Fig. 1**) that allows statistical queries on a set of data while protecting the privacy of individual data points. It works by systematically introducing randomness into query results so that a particular input cannot be reliably evaluated, even by an adversary with external information. For example, a study (Mohammed et al. 2011) demonstrated that participants in a private genomic database could be re-identified by cross-referencing DNA sequences with just 75 additional demographic attributes. Differential privacy counters such linkage attacks by bounding the influence of any single record on output distributions. Formally, a randomized algorithm satisfies ε-differential privacy if for any two input datasets D1 and D2 differing by only one element, the chance of a particular output changes by no more than a factor of exp(ε). Typical values of privacy loss parameter ε range from 0.1 to 10 depending on application sensitivity. Lower epsilon values enforce stricter indistinguishability by mandating higher noise levels.

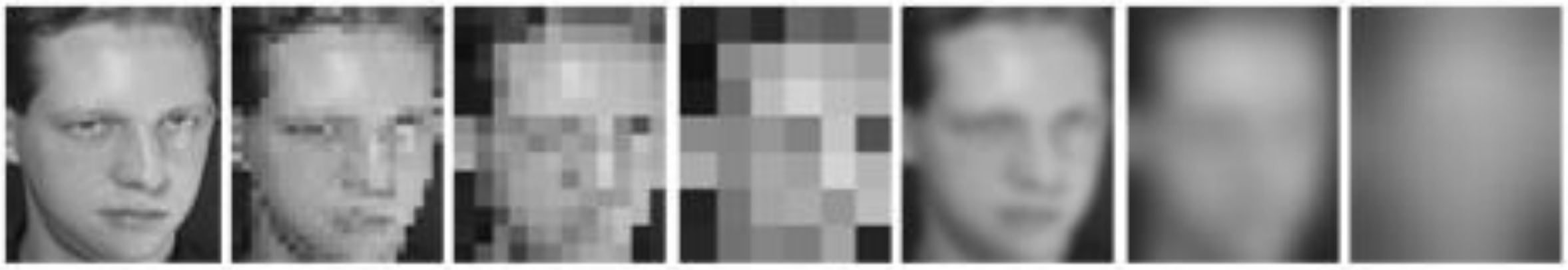

**(a) orig (b) b=4 (c) b=8 (d) b=16 (e) k=15 (f) k=45 (g) k=99**

Fig. 1. A sample of the image (Xue 2021).

*2) Applications in Image Processing*

Image processing pipelines perform various analytical tasks on visual datasets - classification, object detection, image restoration, etc. However, the wealth of sensitive information encoded in images means that unintended leakage from these computations risks exposing confidential attributes. For instance, a study (Fredrikson et al. 2015) showed that facial recognition models can reveal the participation of specific individuals in training datasets with >90% accuracy through output analysis. Such privacy violations severely undermine trust. Differentially private mechanisms bound the impact of individual inputs. As an example, a study (Erlingsson et al. 2014) deployed a differential private web recommendation system at Apple which maintained utilization while providing quantifiable privacy protection.

*3) Some other key image-processing applications*

Medical Imaging Diagnostics: Tumor detection algorithms were shown to leak patients' cancer status through model outputs (Song et al. 2021). Differential privacy protects diagnoses while allowing research.

Surveillance Analytics: Queries about aggregated crowd flows and density patterns must not reveal identifiable information (Zhang et al. 2020). DP enables privacy-safe analytics. Overall, differential privacy constitutes a vital component enabling privacy protection for image processing systems.

*4) Technical Implementation*

There are two main ways to use different privacy settings:

1. Input Perturbation: Adding noise to the input prevents the output from being too dependent on any one record. According to (He et al. 2023). Laplace noise with parameter $\alpha$ for $\varepsilon$-difference privacy with $\alpha \geq \Delta f/\varepsilon$ and $\Delta f$ is an application-specific sensitivity metric.

2. Output perturbation: Adding noise to the final outputs according to the new deep learning algorithm developed by (Zhu et al. 2022; Ziller et al. 2021) and their results. The teacher-student training model inherently introduces emotional boundaries.

The careful measurement of noise levels allows them to be used while reaching the privacy target._Hardware accelerators as proposed (Adesuyi & Kim 2019) also improve efficiency. Therefore, continued research is vital to further mature differential privacy techniques for reliably safeguarding images while retaining analytic accuracy. Careful tuning guided by formal privacy frameworks allows customizable deployment across diverse applications.

TABLE I

COMPARISON OF THE ADVANCED PRIVACY-PRESERVING TECHNIQUES

| Privacy-Preserving Technique | Description | Applications in Image Processing | Technical Implementation |
|---|---|---|---|
| **Differential Privacy** | Cryptographic technique allows statistical queries on a dataset while protecting the privacy of individual data points. Introduces randomness into query results to prevent reliable inference of specific entries. | - Image processing pipelines (classification, object detection).<br>- Medical Imaging Diagnostics (Tumor detection).<br>- Surveillance Analytics (Aggregated crowd flows). | - **Input Perturbation:** Adding Laplace noise to inputs to bound sensitivity.<br>- **Output Perturbation:** Adding noise to final outputs through techniques like teacher-student training.<br>- Calibrating noise levels for utility and target privacy levels.<br><br>- Hardware accelerators for efficiency (Adesuyi & Kim 2019)<br><br>- Continued research for maturity and customization.<br>- Tuning guided by formal privacy frameworks. |
| **Secure Multiparty Computation** | Cryptographic solution enabling collaborative analytics without exposing any one party's private data. Allows different entities to run joint computations on collective datasets. | - Privacy-preserving image classification (Erlingsson et al., 2014).<br>- Secure analysis in medical imaging pipelines.<br>- Mitigating supply chain risks in manufacturing. | - Verifiable Secret Sharing.<br>- Oblivious Transfer.<br>- Zero Knowledge Proofs.<br>- Protocols securing against dishonest majority.<br>- Careful structured sequences of interactions between servers.<br>- Advanced cryptographic building blocks (Verifiable Secret Sharing, Oblivious Transfer, Zero Knowledge Proofs).<br>- Enables secure computations over private data.<br>- Used in medical imaging analytics and secure manufacturing analytics.<br>- Continued advancements in computational performance. |
| **Homomorphic Encryption** | Enables computations directly on encrypted data without decryption. Protects confidentiality end-to-end. Allows | - Medical Imaging Analytics.<br>- Biometric Identification Systems.<br>- Surveillance Analytics. | - Fully Homomorphic Encryption (FHE).<br>- Partially Homomorphic Encryption (FPE).<br>- Format-Preserving Encryption (FPE).<br>- Application in medical imaging analytics, biometric identification, and surveillance analytics. |
| | tasks like classification and reconstruction on encrypted images. | | - Challenges include computational overheads.<br>- Hybrid methodologies (HE, MPC, trusted hardware) for efficiency and verifiable privacy.<br>- Ongoing research for scalability and adoption. |
| **Anonymization Methods** | Techniques to remove personally identifiable elements from images, improving privacy. Includes blurring, pixelation, cropping, and masking. Also involves abstract feature representation and learning algorithms using sanitized derivative data. | - Facial recognition masking.<br>- Blurring, pixelation, cropping.<br>- Adding noise to images. | - Model-based frameworks for optimal anonymization.<br>- Abstract feature sets representation.<br>- Quantifying privacy risks for precise editing.<br>- Challenges include finding optimal representations and maintaining utility.<br>- Combining various techniques for robust anonymization.<br>- Balancing privacy with analytic value.<br>- Continued research in efficient anonymization. |

### *B. Enabling Privacy-Preserving Image Analysis Through Secure Multiparty Computation*

The exponential growth of image data from a plethora of sensors and imaging tools has opened huge opportunities for usable insights through image analysis medical diagnostics, autonomous systems, technologies that are automated, and observational studies are just some of the areas of use that also show strong results. Individually identifiable information, private intellectual property, and infrastructure vulnerabilities are documented deeply at pixel levels and analyzed with vision algorithms The risk of inadvertent data leaks continues to grow as exploration is advanced through deeper, more powerful roots.

### *C. Secure Multiparty Computation*

The Secure Multiparty Computing (MPC) protocol is a cryptographic solution that bridges the separation between high-use sensitive image data and protects privacy MPC allows different companies to collaboratively analyze the collected datasets, without revealing the private data of one party to others. Even the published consolidated outputs do not contain more heuristic information than is strictly necessary to compute the target algorithm. Assessment. The strong assurance model prevents addressing the gap between aspirations of practical intelligence and the realities of conflicting incentives or ownership constraints. Increased access to image data has opened opportunities for insights through image analysis in areas such as health (Fig. 2) and manufacturing (Newton 2017). However, the risks of unintended data leaks from advanced analytics technologies continue to grow. Deep neurons memorize sensory information encoded at the pixel level, as demonstrated in (Fredrickson et al. 2015) On model inversion attacks featuring training data. MPC systems enable collaborative analysis of collected data without revealing confidential information from any one party (Yao, 1982; Goldreich et al., 1987). For example, a study (Nandakumar et al. 2019) used MPC to classify privacy-preservation images across hospitals. Even in the consolidated outputs, there is no predictable information beyond what is needed to compute the target algorithm.

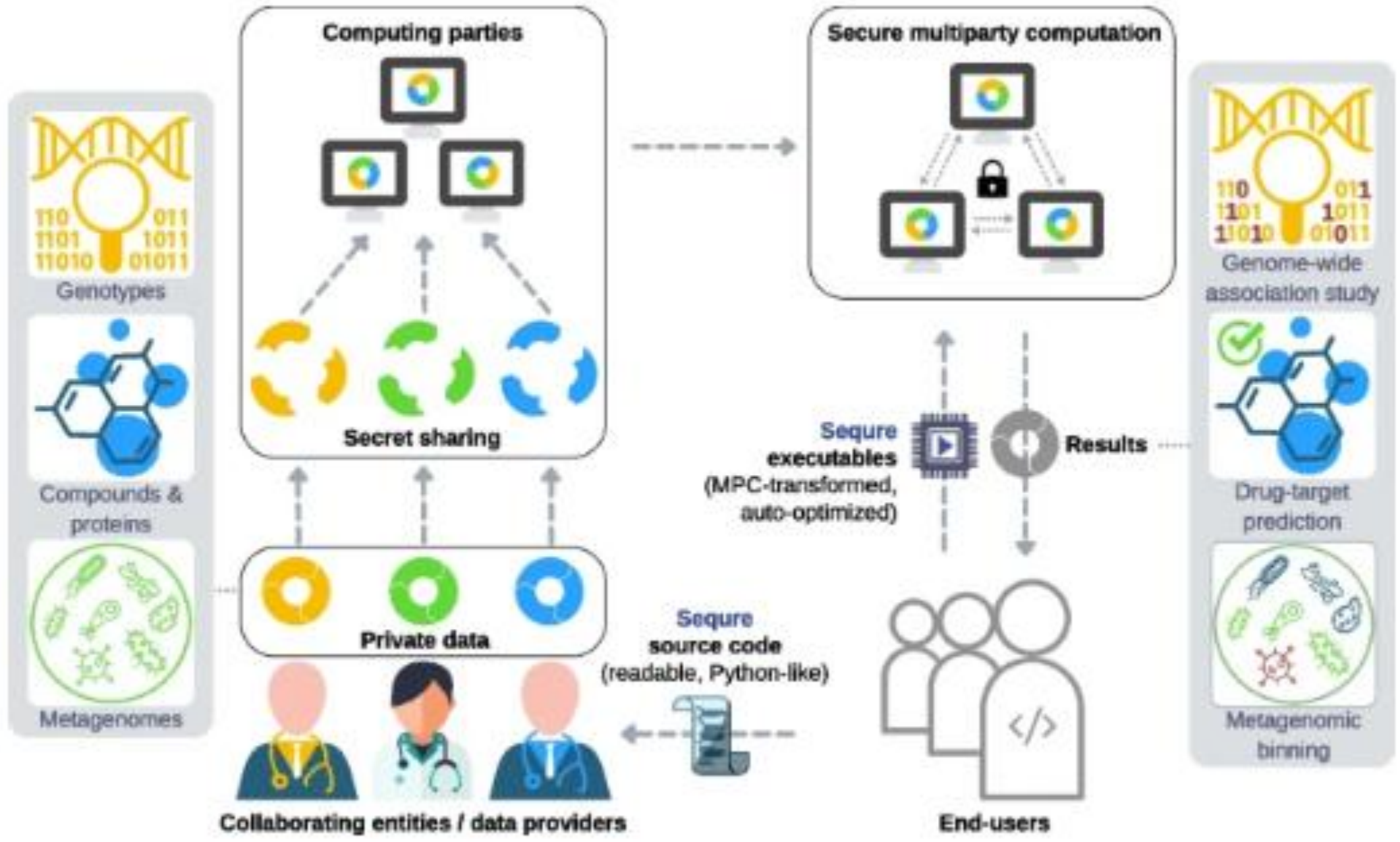

Fig. 2. Secure multiparty computation (MPC) enables collaborative analysis of sensitive private data (Smajlović et al. 2023).

### D. *Foundations of Secure Multiparty Computation*

MPC allows companies to jointly account for usage, through a transparent trust model, by keeping inputs private. Protocols protected against multiple dishonesty limit the damage probability $t < n/2$ sharing servers in a pool of size n (Damgård et al. 2012). At its core, MPC allows multiple companies to calculate performance on their aggregate investments, while keeping investments private. Powerful evidence that limits the statistical results to accurate statistics represents a strong security commitment by MPC against adversaries attempting to steal privacy. There are trust models based on assumptions about enemy boundaries. Protocols protected against majority dishonesty provide strict privacy limiting the damage from the shared server to $t < n/2$ in a pool of size n. A well-ordered sequence of communications between servers efficiently satisfies the objective algorithm without revealing the true privacy values even at intermediate stages on malicious nodes MPC systems use advanced cryptography building blocks.

- Verifiable shared secrets: Sensitive information is loaded into statistically partitioned shares before analysis so that individual shares provide no information about the actual secret. Only authorized quorums that accumulate sufficient minimum shares can reconstruct the encrypted data. Data gets split into shares providing no individual information about actual secrets, as implemented efficiently (Shamir 1979).
- Oblivious Transfer: Enables unconditionally secure transfer of one among many encrypted messages without the sender learning anything about which specific message was obtained. This allows private inputs and outputs during computations. Enables unconditionally secure transfer of one among encrypted messages without the sender learning which message was obtained (Rabin 1981). Together these facilitate practical secure computations over private data.
- Zero Knowledge Proofs: Users can definitively prove protocol adherence without revealing underlying private data. This prevents fake adherence claims aimed at data smuggling.

Together these mechanisms facilitate practical secure computations, unlocking tremendous collaborative potential in analytics over confidential data.

### 1) *Secure Analysis in Medical Imaging Pipelines*

Medical imaging pipelines generating diagnosis inputs through CT scans, pathology slides, MRI images, etc. rely extensively on analytics for anomaly detection, quality measurement, modality translations, and identifying imaging biomarkers indicative of disease prognosis. However, limitations around sharing sensitive patient data continue to obstruct research progress by restricting datasets - especially rare conditions - for more robust model development. Analysis for medical imaging relies on sensitive patient data, which hampers research progress due to established limitations. MPC overcomes such obstacles (Rivest et al. 1978):

1. Sites visual data and see answers in mass training deep learning models.

2. Secure matrix multiplication enables joint scores (Mohassel & Zhang 2017).

3. Only the final collected sample is displayed with the raw medical images displayed. Such a solution accelerates the analysis of joint data as successfully implemented (Vulapula & Srinivas 2018).

MPC overcomes such limitations by using large, distributed learning algorithms in image analysis:

1. Local websites use connectively homogeneous cryptography schemes to encrypt scans and other sensitive visual data using connectively homogeneous cryptography schemes before sharing them with organizations.

2. The combined scores on the mass-assembled data sets are then used to train search models using protocols for secure matrix multiplication and make corresponding comparisons with deep neural network architectures.

3. Propagating gradient updates in a distributed class preserves confidentiality while restoring global model performance.

4. Only the final combined sample is displayed for analytical use without a random display of individual patient clinical images.

Such solutions extend to larger medical organizations on a desirable basis to accelerate research and overcome data limitations that have hampered the development of diagnostic tools of social importance.

### 2) *Mitigating Supply Chain Risks in Manufacturing*

Modern distributed manufacturing relies on scope, sensors, vision tracking systems, and images to maintain quality parameters, business communication, product development, etc. However, manufacturers worry about competition, audit compliance, or IP protection. MPC opens redundant analytics such as error analysis and predictive maintenance in enterprise workflow outsourcing without the need for sensitive data centrally:

1. Assembly line images are still secure in the factory building while only encrypted meta-data is shared with analytics providers through the earlier MPC stages.

2. Collaborative computing on such distributed data enables critical diagnostics - equipment anomalies, production efficiency metrics, etc. - to validate those necessary for optimizing processes.

3. MPC techniques using secure collections in deep learning can address vulnerabilities without allowing random image inputs to be detected.

Overall, MPC opens the way for scalable privacy-preserving intelligence in myriad image analysis areas of private and competitive data Previously security operations were impossible because conflicting priorities became possible by inducing incentives through cryptographic assurance. As the computational performance of MPC protocols becomes more advanced, such solutions will expand analytics ecosystems that rely on sensitive visual data.

## III. PRIVACY-PRESERVING MACHINE LEARNING MODELS FOR IMAGE PROCESSING

As extensive data development continues to drive rapid advances in computer vision capabilities, image analysis models have become increasingly embedded in critical areas such as healthcare, finance, infrastructure security, etc. However, standard deep learning pipelines reveal serious privacy risks - memorizing rails data, allowing reverse attacks, enabling unauthorized profiling e.g. It was shown to take place through the mesh (Fredrikson et al. 2015). Such vulnerabilities hinder adoption in important privacy-critical AI applications. Fortifying the machine learning algorithms themselves with privacy-enhancing techniques provides strong security. Cryptographic techniques such as Homomorphic encryption and differential privacy may prevent accidental memorization and misuse of sensitive training data. Algorithms specifically designed to operate on encrypted data or modify internal representations also limit explicit replication attacks. The weak reward classes in ML models both in training and theory are bridged.

## IV. DIFFERENTIAL PRIVACY GUARDS AGAINST INFERENCES

Strict techniques of differential privacy in algorithms that restrain the intrusion of individual data sets by adding random noise into model production during the estimation Phase safety definition is discussed concerning whether the output distribution will not change drastically on the addition of any single input due to theirs being removed. Reduced noise injection attenuates remembering but maintains the innate patterns. DP allows for cryptographically verifiable anonymization, which is well suited to highly sensitive medical data such as magnetic resonance and other imagery or biometrical information. With an emphasis on the utility, however, hyperparameter tuning is required to be more cautious.

### *A. Federated Learning Allows Decentralized Modeling*

Sensitive data, in this case, participant or silo involved, are distributed across different machines; thus, federated learning can be defined as a program of distributing other equipment for developing any model. The paradigms that are locally trained are very well-bounded within the big fields of global practices by companies such as Google and Hospitals (Brisimi et al. 2018). Call for a Multidimensional Approach of the Study, Reflecting Aly Addressed Mixes of CC Effectiveness, Incentives Use and Privacy as Dynamic Research Frontiers (Pap Ebendo et al. 2019; Carrose et al. 2021) In summary, federated protocols lower the risks involved with centralized data collection by ensuring privacy circulation.

#### *1) Homomorphic Encryption Permits Encrypted Computation*

In this case, the homomorphic encryption schemes can proceed to calculate directly on ciphertexts without converting them into a plaintext format in the process. That makes it possible to run the models of neural networks while preserving end-to-end secrecy. The first-ever conceptualization of homogeneous model estimation is depicted in Crypto Nets by Gilad Bachrach et al. Applications more recently (Kumar et al. 2020) and SEALion (Tim 2019) have achieved ~3x overhead encrypted deep learning for wide models. Approaches within this hybrid towards using HE, MPC, and reliable hardware seem on the right track for effective and verifiable privacy. As facilitate systems to use simple image analytics for meaningful analysis with strong privacy promises, these leading machine learning algorithms ensure that continuous analysis will enforce accuracy and performance effectiveness reinforcement, which encourage implementation in practice. Generally, they are irreplaceable means enabling both privacy and enhanced value to be accomplished simultaneously through visual data analytics across domains.

### *B. Homomorphic Encryption in Protecting Image Data*

Homomorphic encryption allows computations to be performed on encrypted data without the need for decryption everywhere. In image analysis, Homomorphic encryption schemes enable a variety of processing operations - classification, classification, reconstruction, etc. - to be performed directly on an encrypted image, generating encrypted output. This protects the confidentiality of original images as well as derived visual information. Fully homomorphic encryption (FHE) permits arbitrary computations on ciphertexts. Partially homomorphic encryption allows a subset of operations - for instance, multiplication or addition alone. Format-preserving encryption (FPE) retains data formats after encryption.

*Homomorphic encryption has powerful applications in privacy-preserving image analysis:*

- Medical Imaging Analytics: Algorithms can identify anomalies, tumors, etc. in encrypted diagnostic images without exposing patients' confidential health information.
- Biometric Identification Systems: Facial recognition, fingerprint matching, etc. can be performed on encrypted probe images to verify identity while securing biometrics.
- Surveillance Analytics: Queries like counting people or tracking objects can run on encrypted video feeds without revealing identities.

Role of encryption in image data security: Encryption is key to protecting the privacy of image data when stored and transmitted over networks Both traditional encryption protocols such as AES and RSA and emerging uniform systems provide a strong security foundation for image pipelines. Custom ciphers provide generalized security for image files at rest in applications - such as text documents. However, they must be defined before any work that may interfere with subsequent research projects. Special protocols such as format-preserving encryption have a visual aspect that allows certain types of encrypted domain operations. Homomorphic encryption (Fig 3) makes it easier to perform arbitrary computation on encrypted data and holds promise for privacy-preserving image analysis.

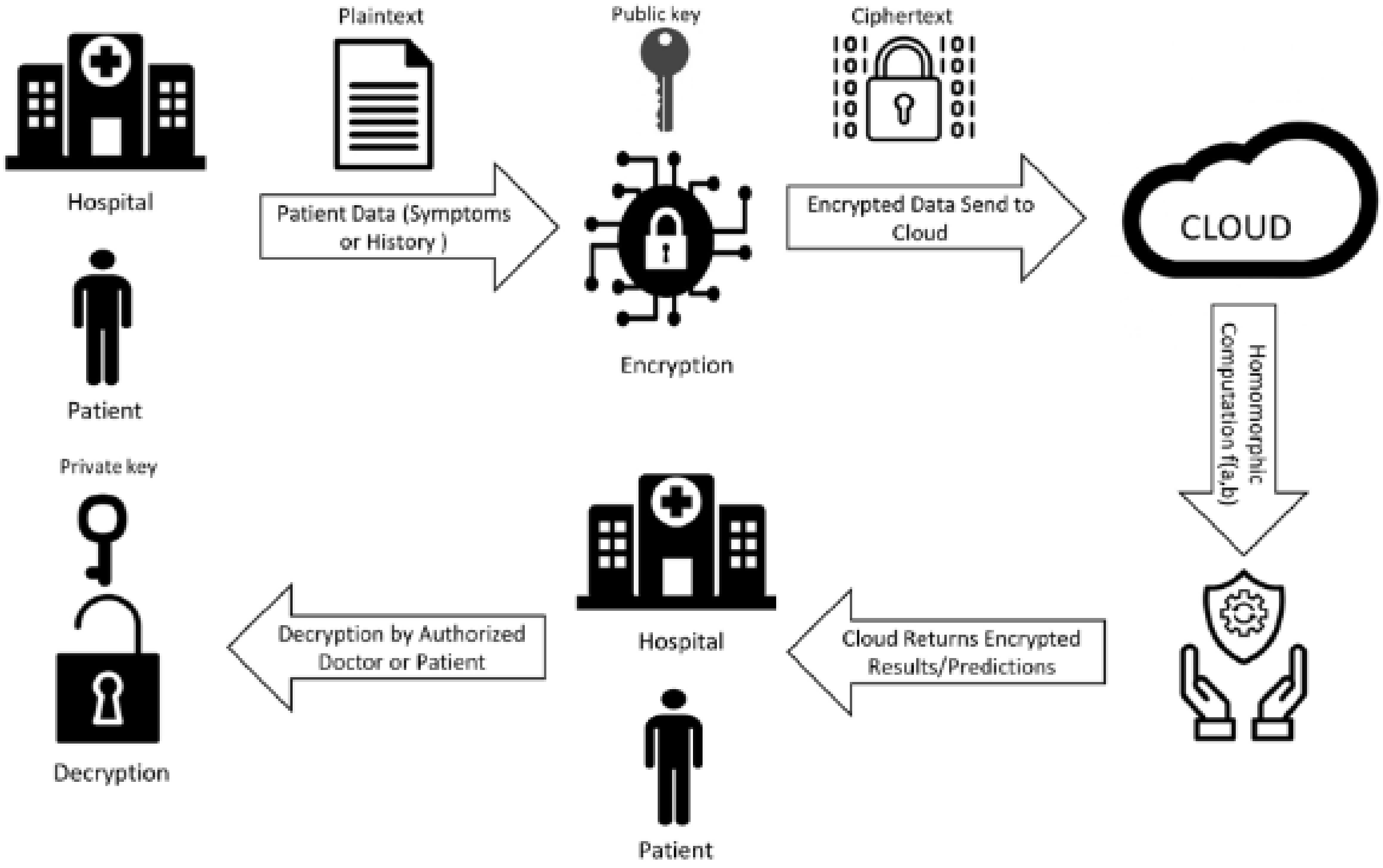

Fig 3. Homomorphic encryption in the healthcare industry (Munjal & Bhatia 2022).

### C. Anonymization Methods for Image Processing

Anonymizing image data, for example, by taking out elements that could identify a person, is a way to enhance privacy directly. These methods include smudging, pixilating, trimming the edges with face covers of licenses and identity cards facing detection placing a mask over such areas. To accomplish the noise, unique biometrics such as fingerprints can also be an accurate method of identification. On the other hand, results of various studies have indicated that several ad hoc manipulations are still inadequate enough to fully anonymize such images, thus enabling one to reconstruct original feelings. C-forms of strong anonymized models need tailored-made models uniquely measuring the risk level of private information within a given data set and addressing factors that significantly makeup user distinction. Model-founded rationale, therefore, is more practical than ad hoc approaches. As an alternative, the output is for a given of some abstract objects that represent an image rather than the use of actual sensitive pixel values. An example is the encoding of only meta-level scene attributes instead of the whole video scene. The learning algorithm can then create a clean source from which it is possible to replace the original images. Such a measure eliminates the purposed disclosures and minimizes the privacy implications of data breaches. It then poses a question on how an optimal representation can be constructed so that there is minimization of privacy risk while at the same time maintaining the utility. On average, the infusion of image processing strategies with quantized models and abstract feature sets produces robust anonymity which stands as a compromise between privacy needs and analytical benefits.

### D. Balancing Confidentiality with Utility in Image Processing

The challenge in privacy-preserving image analysis is the trade-off between preserving information so that research still can generate useful results and at the same time ensuring privacy. Although the utilization of a relatively large scale may infringe on several vital structural elements and greatly harm some very valuable information. On the other hand, weak anonymity creates data that is easily infiltrated through private ones. Ideally, privacy options would ensure that only those accounting processes required or having the potential to facilitate disclosure for the proper functioning of the application are permitted. Full homogeneous encryption is the closest but involves useless overheads. In practice, current methods impose tolerances on the selection of supported operations, product accuracy, or efficiency. Balancing this trade-off requires that the system be designed to limit the minimum distortion of only irrelevant statistics. Crypto-assisted machine learning constructs that train complex patterns on encrypted data provide greater flexibility. Another approach is to solve optimization problems of achieving maximum accuracy with transformed inputs. Despite potential improvements that expand the scope of data manipulation, robust systems are also potentially future-proof privacy. They target the main causal mechanisms behind private tables rather than surface patterns. Overall, privacy and usability require only an understanding of application-specific information to judiciously introduce uncertainty where it preserves privacy without negative impact.

### E. Ensuring Accuracy in Privacy-Preserving Image Techniques

It masks sensitive features in images, and confidentiality-preserving transformations can hinder the accuracy of the analysis and modeling performance if adequate precautions are not taken Methods contribute to accuracy even when privacy is used as mentioned below:

• Determine the accuracy/privacy trade-off under different standards to tune the best settings in each application.

• Layer privacy protocols allow partial protection rather than full visits.

• Configure computer vision models to increase focus on intangible objects.

• Preliminary training data such as expected distortion data will be available at runtime.

Overall, the balance between information protection and privacy remains contextual. However, following principled approaches enhances the accuracy achieved under privacy constraints by using accuracy-focused algorithms and AI adapted to non-continuous data.

### *F. Protecting Individual Privacy in Image-Intensive Fields*

Many emerging applications such as social networks and autonomous systems involve capturing, storing, and analyzing images on a massive scale - the greater the personal privacy risks, the more they are created in the form of images posted on social platforms research to identify users, target ads, and predict trends without approval. Features such as auto-alt text highlighting the identity reflect a reduction in default security. Enhanced privacy requires initial efforts across all technologies, not just enforcement boxes. With driver-assist technologies, intelligent urban systems, and even domestic robots becoming more ubiquitous, these environmental devices storing moving images and sensory data expanded data requests pose unprecedented risks to privacy through behavior modeling, detection, and tracking.

## V. FUTURE DIRECTIONS

As rapid development continues to expand the capabilities and use cases of visual data, continuous improvements in privacy protection techniques are needed. Several promising strategies remain for transforming privacy management practices to perform image analysis. Providing classical- Crypto primitives that match image type rather than generic data that compromises ciphers also improves performance. Integrated learning allows joint model training across organizations without exposing local image data sets. Emerging strategies further reinforce privacy in cross-silo learning. New homologous encryption variants reduce overhead by representing images compactly and optimizing expensive operations such as bootstrapping. Dedicated hardware such as GPUs, TPUs, and Neural Engines will accommodate them. Drawing on insights into law and regulation, the privacy framework in place will help to empirically assess risks, develop appropriate best practices, and guide responsible data processing governance Therefore, continued research combining advanced techniques from cryptography, AI/ML, and theoretical computer science meets the expanding demands of future imaging. Doable and ready to engineer privacy solutions Careful participatory planning can also motivate the active development of ethical visual intelligence systems.

## VI. CONCLUSION

Rapidly evolving image processing systems have transformed various aspects of our lives, from healthcare to surveillance to social media. However, this shifting dynamic adds a significant puzzle: the privacy challenges posed by the massive collection, storage, and processing of image data grow as we navigate this challenging terrain. The ubiquitous-nature of introductory search lays the groundwork for understanding the complex web of privacy concerns that shroud this technical field. Important information contained in images, including appearance, behavior, associations, locations, and emotional state types highlight the magnitude of potential privacy breaches (Table 4). Not only is it a breach of privacy but it represents an important ethical concern that needs immediate attention.

TABLE II

COMPARISON OF PRIVACY-PRESERVING TECHNIQUES

| **Criteria** | **Differential Privacy** | **Secure Multiparty Computation (MPC)** | **Homomorphic Encryption** | **Anonymization Methods** |
|---|---|---|---|---|
| **Principle** | Introduces randomness into query results to protect individual data points. | Allows collaborative analytics without exposing private inputs. | Enables computations on encrypted data without decryption. | Removes personally identifiable elements to improve privacy. |
| **Applications** | - Image processing pipelines (classification, object detection).<br>- Medical Imaging Diagnostics.<br>- Surveillance Analytics. | - Privacy-preserving image classification across hospitals.<br>- Secure analysis in medical imaging pipelines.<br>- Mitigating supply chain risks in manufacturing. | - Medical Imaging Analytics.<br>- Biometric Identification Systems.<br>- Surveillance Analytics. | - Facial recognition masking.<br>- Blurring, pixelation, cropping.<br>- Adding noise to hinder recognition. |
| **Implementation Techniques** | - Input Perturbation: Adding noise to inputs.<br>- Output Perturbation: Adding noise to final outputs. | - Verifiable Secret Sharing.<br>- Oblivious Transfer.<br>- Zero Knowledge Proofs. | - Fully Homomorphic Encryption (FHE).<br>- Partially Homomorphic Encryption. | - Blurring, pixelation, cropping, masking.<br>- Adding noise to images.<br>- Abstract feature representation. |
| | | | - Format-Preserving Encryption (FPE). | |
| **Challenges** | - Tuning privacy loss parameter ε for a balance between privacy and utility.<br>- Computational overheads can affect utility. | - Coordination among multiple parties.<br>- Overheads associated with secure computation. | - Significant computational overheads.<br>- Limited support for complex operations in FHE. | - Adequate anonymization without compromising utility.<br>- Quantifying privacy risks for precise editing. |
| **Advantages** | - Rigorous statistical formulation.<br>- Bounds leakage of individual data records. | - Enables collaboration without exposing raw data.<br>- Preserves privacy during joint computations. | - Allows computations on encrypted data without decryption.<br>- Protects confidentiality end-to-end. | - Direct removal of personally identifiable elements.<br>- Can be tailored for specific privacy risks. |
| **Use Cases** | - Image classification, object detection.<br>- Medical imaging diagnostics.<br>- Surveillance analytics. | - Collaborative analytics in medical imaging.<br>- Privacy-preserving image classification.<br>- Secure manufacturing analytics. | - Medical imaging analytics.<br>- Biometric identification systems.<br>- Surveillance analytics. | - General image anonymization for privacy improvement. |
| **Future Directions** | - Research in optimizing ε for different applications.<br>- Advancements in accuracy and efficiency. | - Improved communication efficiency in federated learning.<br>- Integration with emerging technologies. | - Reduction of computational overheads.<br>- Integration with trusted hardware.<br>- Application-specific optimizations. | - Development of model-based frameworks for optimal anonymization.<br>- Exploration of abstract feature sets. |

One of the key pillars of privacy challenges in image processing resides in the realm of surveillance and healthcare. Surveillance systems, omnipresent in our modern world, raise concerns about civil liberties and unwarranted tracking. The paper advocates for privacy-preserving techniques such as access controls, anonymization of faces, and secure multi-party computation to counterbalance the intrusive nature of surveillance. Similarly, in healthcare, where medical images contain highly sensitive patient information, the importance of techniques like homomorphic encryption and differential privacy cannot be overstated. It requires everyone to find innovative solutions that not only provide the needed quality health issues but also mind patients' privacy, which calls for an approach that involves skill as well as art at the same time.

Moreover, the emphasis on state-of-art privacy-preserving strategies and specifically discussion of the density of differential privacy in image processing reveals the inner details of safeguarding personal data under forces of analytical means. The discussion highlights the utility of randomized algorithms in applying uniformly noise through a systematic approach, thus obtaining a strong defense against model inversion attacks and other such privacy threats. Confidentiality issue creates an important obstacle in preventing irresponsible flow from the imaging estimates, thereby assuring, and maintaining trust and confidence in the risk management needed to apply advanced analytical models.

Secure Multiparty Computing (MPC)procedure supplies a cryptographic lifespan to fill up the gap between tips defogging useful knowledge from confidential image data, and maintaining privacy MPC provides an innovating answer to privacy issues with losing confidentiality transparently and enabling collaborative research medical image pipelines provide the ability of MPC to perform rapid analysis and retain privacy for patients, while varieties show its efficiency. Privacy-preserving machine-learning models discover weaknesses in standard studies of deep pipes and cryptography like homomorphic privacy, as well as the concurrency form indicators, unlike traditional overlaid Turing schemes. These tactics together with government learning can rather be considered as alternative approaches to manage risks related to large-scale collection. What is noteworthy in the discussion concerning future research directions is that it promotes accuracy and efficiency but will result in a well-grounded practice of enhanced privacy.

The Secure Multiparty Computing (MPC) protocol provides a cryptographic lifecycle to bridge the separation between extracting valuable insights from sensitive image data and protecting privacy MPC offers a transformative solution to privacy challenges by transparent confidentiality and enabling collaborative research. Applications to medical image pipelines highlight the ability of MPC to perform rapid analysis while preserving patient privacy and demonstrate its versatility Exploring privacy-preserving machine-learning models look for vulnerability types of standard studies of deep pipelines and cryptography such as Homomorphic privacy, and unique privacy. Introduces strategies. These strategies combined with government learning represent alternative ways to mitigate the risks associated with centralized data collection. The discussion highlights the need for continued research to promote accuracy and efficiency and will lead to the adoption of privacy-enhancing strategies in practice. Homomorphic encryption appears as a beacon of hope for protecting image data, allowing computation to be performed on encrypted data without compromising privacy. Acknowledging the challenges presented by cybernetics, homogeneous encryption is crucial for extracting valuable image analysis insights without sacrificing privacy Its applications in medical imaging, biometric identification systems, and surveillance analytics demonstrate its potential demonstrated in various industries. Anonymization techniques ensure consistency in the usability of privacy-preserving copy techniques and the benefits of privacy Build the small strategies needed to jointly overcome complex privacy challenges. Emphasis is an ethical visual intelligence focused on protecting personal privacy in image-intensive environments including social networks and enabling systems. Ring Emphasizes the need to create try and emphasize participation in planning.

As the paper concludes, promising future directions for privacy protection for image processing are identified. Quantum-resilient encryption protocols, federated learning, new isotropic encryption variants, and dedicated hardware stand out as potential game-changers Combining advanced technologies in cryptography, AI/ML, and theoretical computer science with privacy settings date design and design considerations towards robust and scalable privacy solutions Hold the technical key. Essentially, while the paper addresses privacy challenges in imaging processing, a holistic effort together. It is necessary to move forward which requires technological innovation, ethical considerations, and policy frameworks to ensure the fundamental right to privacy in the digital age is respected, with transformative power, the fundamental right to privacy plays a digital age.